\documentclass[allclo,superscriptaddress,eqsecnum,amsfonts,showpacs]{revtex4}
\usepackage{epsfig}

\newcommand{\be}{\begin{equation}}
\newcommand{\ee}{\end{equation}}
\newcommand{\bea}{\begin{eqnarray}}
\newcommand{\eea}{\end{eqnarray}}

\newcommand{\ep}{i\varepsilon}
\newcommand{\nn}{\nonumber}

\begin{document}

\preprint{ \parbox{1.5in}{\leftline{hep-th/??????}}}

\title{Confinement, Chiral Symmetry Breaking and complex mass in QED2+1 in Minkowski and Euclidean spaces}

\author{Vladimir ~\v{S}auli}
\affiliation{CFTP and Dept. of Phys.,
IST, Av. Rovisco Pais, 1049-001 Lisbon,
Portugal }
\affiliation{Dept. of Theor. Phys., INP, \v{R}e\v{z} near Prague, AV\v{C}R}

\author{ Zoltan Batiz}
\affiliation{CFTP and Dept. of Phys.,
IST, Av. Rovisco Pais, 1049-001 Lisbon,
Portugal }

\begin{abstract}
Without any analytical assumption we solve the ladder QED2+1 in Minkowski space.  
Further, we transform Greens functions  
to the Temporal Euclidean space, wherein we  show that in the special case of ladder 
QED2+1 the solution  is fully equivalent to the  Minkowski one.
 The obtained complex fermion propagator exhibits confinement because of the lack of the  
pole at the real timelike $q^2$ axis.

\end{abstract}

\maketitle
%

\section{Introduction}

 Quark confinement in Quantum Chromodynamics (QCD) is a phenomenon of current
interest. Due to the fact that QCD is not easily tractable, the toy models which exhibit QCD low energy phenomena --the confinement and chiral symmetry breaking-- are often investigated.
In this respect 2+1d Quantum Electrodynamics (QED2+1) has these similarities with QCD in the
usual 3+1d Minkowski space.  
Based on the Euclidean space study of QED3, the chiral symmetry breaking for a small number flavors has been proposed  for the first time in \cite{APNAWI1988}. Since the scale of dynamical chiral symmetry breaking, being characterized  by a fermion mass in the infrared -$M(0)$-, is
one order of magnitude smaller then the topological dimensioned coupling $e^2$, the 
Schwinger-Dyson equations (SDEs) provide a unique powerful framework for the nonperturbative study, see e.g. most recent studies \cite{BARA2007,BARACORO2008}. 
The importance of the unquenching effect in QED3 for an increasing $N_f$ has been recognized a long time ago \cite{DAKOKO1989,BUPRPRO1992} and reinvestigated in SDEs framework lately \cite{GUHARE1996,MARIS1995,FIALDAMA2004,BARA2007}. Particularly, confinement in relation with  dynamical  complex pole generation in fermion propagator has been discussed in 
\cite{MARIS1995}.  Further analogy with QCD in finite temperature and chemical potential has been explored   
\cite{HEFESUZO2007}. It is noteworthy that QED3 is of current interest as an effective theory of high-temperature superconductors \cite{FRTEVA2002,HERBUT2002,ASTE2005,THHA2007} , Mott insulator \cite{NOHA2005} and graphene \cite{NOVO2005,GUSHCA2007}.

However, the all aforementioned  studies have been done in the
standard Euclidean
space. That is, after performing the standard Wick rotation \cite{WICK} of the timelike
components of the momentum variables (internal integral momentum as well as external one, explicitly $p_3^{E}=-ip_0^{M}$,  the measure $id^3p^{M}=-d^3p^{E}$). It is assumed and widely believed, that the Green`s functions for timelike arguments can be obtained after analytical continuation of the functions calculated in  Euclidean space, where it is supposed the Euclidean solution itself  represent the correct Minkowski solution for the spacelike arguments (for some discussion see also   Sections 2.3 and 6.3 in  \cite{ROBERTS}).
Therefore, to shed a new light and for the first time,  we solve fermion SDE directly in 2+1 Minkowski space. The so called ladder approximation of electron SDE is introduced in the the Section II. 

In the  Section III, by using hyperbolic coordinates, we 
solve the SDE directly in the Minkowski space. While for timelike $p^2$ it provides selfconsistent gap equation, i.e. the integral equation with the mass function of the same arguments in both sides of Eq.,   
whilst for spacelike Minkowski subspace the solution is naturally constructed from necessarily  known timelike   solution. Furthermore, we derive the SDE in the Temporal Euclidean ($E_T$) space in the Section IV and prove that this exactly leads to the  original  Minkowski formulation for timelike momenta. Recall here,  $E_T$  space  metric is obtained from Minkowski one by the multidimensional  Wick rotations, but now instead for the time component, it is  made for  all the space coordinates of the Lorentz three vector \cite{SAZO2008}. 

 The numerical results for various ratio of the coupling and the electron mass 
are presented in Section V. The imaginary part of the mass function is automatically generated for a coupling strong enough and resulting fermion propagator has no mass singularity in the timelike region.  Such electron electron can never be on-shell and thus never observed as a free particle, in other words it is permanently confined \cite{CORN1980,GRIBOV91,ROWIKR1992,MARIS1995}.
 We further discuss the (in-)validity of standard Wick rotation and  conclude in the Section VI.

\section{Fermion SDE in QED2+1, QED3 in ladder approximation}

In our study we employ Minkowski metric $g_{\mu\nu}=diag(1,-1,-1)$, in order to  properly describe chiral symmetry, we use the standard four dimensional Dirac matrices such that they anticomutation relation is  $\left\{\gamma_{\mu},\gamma_{\nu}\right\}=2g_{\mu\nu}$. With these conventions the inverse of the full fermion propagator reads
\bea \label{gap}
S^{-1}(p)&=&\not p- m-\Sigma(p)
\nn \\
\Sigma(p)&=&ie^2\int \frac{d^3k}{(2\pi)^3} \, , G^{\mu\nu}(k-p)\Gamma_{\mu}(k,p)S(k)\gamma_{\nu}
\eea

We consider the explicit chiral symmetry breaking  mass term of the form $m\bar{\psi}\psi$
so  parity is conserved. In this case the dressed fermion propagator can be parametrized
by two scalar function like
\be
S(p)=S_v(p)\not p+S_s(p)=\frac{1}{\not p A(p)-B(p)}\, . 
\ee

The full photon propagator $G$ and the electron-positron-photon vertex $\Gamma$ satisfy their own
SDEs (for their general forms see \cite{ROBERTS})   

The ladder approximation is the simplest selfconsistent approximation which approximate  the unknown Greens functions be their free counterpartners, i.e. $\Gamma_{mu}=\gamma_{mu}$ and the photon propagator in linear covariant gauges is
\be
G_{\mu\nu}=\frac{-g_{\mu\nu}+(1-\xi)\frac{k_{\mu}k_{\nu}}{k^2}}{k^2} \, .
\ee

\section{Direct Minkowski space calculation}

In general  QFT the Greens functions are not  real functions but complex tempered distributions.
In perturbation theory these are just real poles (together with its Feynman $\ep $ prescription) of the propagators, which when coincide in the loop integrals, produce branch cut starting at the usual production threshold. At one scalar loop level, the two propagators make the selfenergy complex above  the point $p^2=(M_1+M_2)^2$, wherein $M_1,M_2$ are the real masses- in fact the positions of these poles. Depending on the masses of the interacting fields, the real pole persists   when  situated  below the threshold or we get non-zero width and the free particle becomes resonance with finite lifetime.

 In strong coupling quantum field theory the mechanism of complexification can be very different (however the mixing of both mechanisms is not excluded). 
Here we simply assume that there is no zero at the inverse of propagator for real $p^2$, thus 
$\ep$ factor is not necessary and we integrate over the hyperbolic angles of Minkowski space directly.
For this purpose we have to consider the propagator function as the complex one for all $p^2$.
 A convenient parametrization of the  complex fermion propagator functions $S_s$ and $S_v$  can be written as
\bea
S_s(x)&=&\frac{B(k)}{A^2(k)k^2-B^2(k)}
\nn \\
&=&\frac{R_B\left[(R_A^2-\Gamma_A^2)k^2-R_B^2-\Gamma_B^2\right]+2R_A\Gamma_B\Gamma_A\,k^2}{D}
\nn \\
&+&i\, \frac{\Gamma_B\left[(R_A^2-\Gamma_A^2)k^2+R_B^2+\Gamma_B^2\right]-2R_BR_A\Gamma_A\,k^2}{D} \, ,
\label{ss}
 \\
S_v(k)&=&\frac{A(k)}{A^2(k)k^2-B^2(k)}
\nn \\
&=&\frac{R_A\left[(R_A^2+\Gamma_A^2)k^2-R_B^2+\Gamma_B^2\right]-2R_B\Gamma_A\Gamma_B}{D}
\nn \\
&+&i\, \frac{\Gamma_A\left[-(R_A^2+\Gamma_A^2)k^2
-R_B^2+\Gamma_B^2\right]+2R_A R_B\Gamma_B}{D}\, \, ,
\label{sv}
\eea

where $R_A,R_B$ $(\Gamma_A,\Gamma_B)$ are the real (imaginary) parts of the functions $A,B$
and the denominator $D$ reads
\be
D=([R_A^2-\Gamma_A^2] k^2-[R_B^2-\Gamma_B^2])^2+4(\Gamma_A R_A-\Gamma_B B)^2 \,.
\ee

In order to be able to compare between Minkowski and lately  considered Euclidean  spaces, the transformation in use should leave the Minkowski spacetime interval 
\be
s=t^2-x^2-y^2
\ee
manifestly apparent (i.e. $s$ could be a variable of the integral SDEs).   

To achieve this we will use 2+1 dimensional pseudospherical (hyperbolic) transformation of 
Cartesian Minkowski coordinates.
The obstacles followed by Minkowski hyperbolic angle integrals when going beyond $A=1$ approximation restrict us to the Landau gauge wherein the $A=1$ is the exact result in Temporal Euclidean space.

In momentum space our convenient choice of the substitution is the following:

\bea \label{myway}
\int d^3k K(k,p)=&&\int_0^{\infty} dr r^2 \int_0^{2\pi} d \theta  \int_0^{\infty} d \alpha
\left\{\sinh\alpha \, {\mbox{\begin{tabular}{|c|}
$k_o=-r\, \cosh\alpha $ \\ $k_x=-r \,\sinh\alpha \, \sin\theta$ \\
$k_y=-r \,\sinh\alpha\, \cos\theta$ \\
\end{tabular}}}
+\sinh\alpha\, {\mbox{\begin{tabular}{|c|} $k_o=r\, \cosh\alpha$\\ $k_x=r \,\sinh\alpha\, \sin\theta$\\ $k_y=r \, \sinh\alpha\, \cos\theta$
\end{tabular}}}\right. 
\nn \\
&&+\left.\cosh\alpha\, {\mbox{\begin{tabular}{|c|} $k_o=-r \sinh \alpha$\\$ k_x=-r \cosh\alpha\, \sin\theta$\\$ k_y=-r \cosh\alpha\, \cos\theta$
\end{tabular}}}+
 \cosh\alpha\, {\mbox{\begin{tabular}{|c|} $k_o=r \sinh \alpha$\\$ k_x=r \cosh\alpha\, \sin\theta$\\ $k_y=r \cosh\alpha\, \cos\theta$
\end{tabular}}}\right\} \, K(k,p) \, .
\eea

Notice, the integral boundaries are universal for  all the subregions of Minkowski space,
  the first line corresponds to the integration over the timelike 2+1momentum where we have 
\be
k^2=k_o^2-k_x^2-k_y^2=r^2>0 \, ,
\ee
where the left term corresponds to the negative energy interval $k_0<-\sqrt{k_x^2+k_y^2}$
and the right term corresponds to the positive  $k_0>+\sqrt{k_x^2+k_y^2}$. The second line
stands
 for the spacelike regime of the integration
\be  
k^2=-r^2<0\, ,
\ee
wherein the left term corresponds to the energy component interval  $k_0= (-\sqrt{k_x^2+k_y^2},0)$, while the right term in the second line stands for positive $k_0= (0,\sqrt{k_x^2+k_y^2})$ subspace of the full 2+1 dimensional Minkowski space. Functions $V$ in Rel. (\ref{myway}) represents the integrand of SDE.

The functions $A,B$  are Lorentz scalars, thus they can depend on $p^2$ only. We freely take the simple choice  of timelike external momenta as $p_{\mu}=(p,0,0)$ which leads to the following $\alpha$ integrals
 for the timelike part of internal momenta:

\be \label{andula}
\int_0^{\infty} d\alpha  \frac{\sinh \alpha}{r^2+p^2+2pr\cosh\alpha}+
\int d\alpha  \frac{\sinh \alpha}{r^2+p^2-2pr\cosh\alpha}
=\frac{1}{pr}\ln\left|\frac{r-p}{r+p}\right| \, .
\ee

The contribution stemming from spacelike part of loop momenta gives zero because of negative and positive time volume contributions, although each infinite separately, they cancel each other. Note, the $\theta$ integrals contributes by simple  $2\pi$ prefactor.

Integrating over the angles we can see that at the level of our approximation, the SDE separate for spacelike and timelike regime of the threemomenta. For timelike $p$ we get for the function $B$
\be \label{ET1}
B(p)=m+i(2+\xi)\frac{e^2}{4\pi^2}\int_0^{\infty} dk
\frac{k}{p}\ln\left|\frac{k+p}{k-p}\right|S_s(k) \, ,
\ee
where $\xi$ is a gauge parameter (and where we trivialy changed $r$ of (\ref{andula}) to $k$)
Stressed here, the Eq. (\ref{ET1}) is derived without any requirement of analyticity for the the propagator and the kernel.

For  the external spacelike Lorentz threevector of momenta the $\alpha$ integration over the spacelike regime gives zero. Taking for instance $p_{\mu}=(0,p,0)$ this can be most easily seen by the inspection of the integrals
\bea
&&\int_0^{\infty} d\alpha  \frac{\cosh \alpha}{-r^2+p^2+2pr\cosh\alpha\sin\theta}+
\int d\alpha  \frac{\sinh \alpha}{-r^2+p^2-2pr\cosh\alpha\sin\theta}
\nn \\
&&=-\frac{a}{b}\int_0^{\infty} d\alpha\left(\frac{1}{a+b\cosh\alpha}-\frac{1}{a-b\cosh\alpha}\right)
=0
\label{iclka}
\eea  
where $a=p^2-r^2\, ,b=2pr\sin\theta$. The  integrals in bracket (\ref{iclka}) can be evaluated by using of the following formula:
\bea \label{tabule}
&&\int_0^{\infty} d\alpha\frac{1}{a+b\cosh\alpha}=
\frac{1}{\sqrt{a^2+b^2}}\ln \frac{a+b+\sqrt{a^2-b^2}}{a+b-\sqrt{a^2-b^2}} \,
\nn \\
&&\int_0^{\infty} d\alpha\frac{1}{a+b\sinh\alpha}=
\frac{1}{\sqrt{a^2+b^2}}\ln \frac{a+b+\sqrt{a^2+b^2}}{a+b-\sqrt{a^2+b^2}}
\eea
 
The remaining contributions stem from the combination of the external spacelike  and the internal spacelike momenta, putting  now $a=p^2+r^2\, ,b=2pr\sin\theta$ we can get 
\bea \label{beruska}
&&\int_0^{\infty} d\alpha  \frac{\sinh \alpha}{r^2+p^2+2pr\sinh\alpha\sin\theta}+
\int d\alpha  \frac{\sinh \alpha}{r^2+p^2-2pr\sinh\alpha\sin\theta}
\nn \\
&&=-\frac{a}{b}\int_0^{\infty} d\alpha\left(\frac{1}{a+b\sinh\alpha}-\frac{1}{a-b\sinh\alpha}\right)
=\frac{2}{b}\frac{\ln\left(\sqrt{1+(b/a)^2}-b/a\right)}{\sin\theta \,\sqrt{1+(b/a)^2}}
\eea  
where the definite integrals in the second line are given by second line in Eq. (\ref{tabule}) .

Adding the all together we get for $p^2<0$ (for $\xi=0$ and $A=1$) the following integral expression for the function $B$:
\bea  \label{surprise}
&&B(p^2)=m+i\frac{e^2}{(2\pi)^3}\int_0^{\infty} dr \int_0^{2\pi} d\theta \frac{r}{\sqrt{-p^2}}
\frac{\ln\left(\sqrt{1+z^2}-z\right)}{\sin\theta \,\sqrt{1+z^2}}
\frac{B(r^2)}{r^2-B^2(r^2)} \, ,
\nn \\
&&z=\frac{2\sqrt{-p^2} r \sin\theta}{r^2+p^2}  \, ,
\eea

To conclude, we stress the main difference when compared to the standard  treatment. Here this is the timelike part of Minkowski subspace where the results are most naturally obtained.
Quite opposite to the standard approach where Minkowski solution is constructed by continuation of the Euclidean result, here the solution for the spacelike argument is non-trivialy made from the timelike soluiton which must found as a first. Actually, the solution at timelike domain given by Eq. (\ref{ET1}) is needed in the rhs. of (\ref{surprise}).

The validity of standard Wick rotation is highly speculative topics in the literature. Here we know the solution directly in Minkowski space and  the comparison with the Euclidean solution is straightforward and easy task. 
 Assuming the Wick rotation is valid, we could get the Minkowski solution equal to the Euclidean one at spacelike domain of momenta. We anticipate here and it will be explicitly shown in the next Section that the mass function $B$ is actually complex  for all real timelike $p^2$ at large window of studied parameters $m$ and $e$. Hence the Minkowski solution $B(0)$ being complex, is not the one literally known from the Euclidean studies, where $B_E(0)$ is an always purely real number.  These solutions do not coincide in the Minkowski light-cone (Euclidean zero) and in other words: commonly used strategy based on the analytical continuation of the Euclidean (spacelike) solution  to the timelike axis is wrong and should be abandoned in the case of QED2+1 theory. In our simple model, the "continuation" is not the analytical continuation of necessarily holomorphic function,  but is given by Eq. (\ref{surprise}). This should be  reconsidered in any other specific model -approximation and truncation of SDEs- system.

We do not know whether there is a unique general rule to continue between spaces (we mean a simple prescription, which do not require
 the solution of SDE for completely general complex arguments). However, before presenting the numerical solution of the Eq. (\ref{ET1}), we will exhibit the equivalence of QED3 defined in the $ET$ space, where the metric simplifies and allows to consider more complicated approximations of SDE.

\section{Proof of an equivalence of  QED fermion SDE formulated in Minkowski and Temporal Euclidean space.}

In this Section we will show  that the Temporal Euclidean formulation of our considered SDE exactly agrees with the one derived  in Minkowski space.

In the paper \cite{SAZO2008} it was proposed that N-dimensional analog of Wick rotations
performed for space components of Minkowski $N+1$-vector can be partially useful for  the study of a strong coupling quantum field theory.
In even dimensional  3+1QCD it is  the nonperturbative mechanism responsible for complex mass generation which is responsible for the absence of real pole type singularities in Greens functions evaluated at real their arguments which thus makes the nonperturbative calculations feasible there. 

In odd dimensional theory, like QED2+1 we study here, the complexification of masses and couplings can be quite naturally expected  because presence of  $i$  in the measures of the integrals defined in ET space. 
To see this explicitly, let us consider the momenta as complex variables and let us  assume that there are no singularities in the second and the fourth quadrants of complex planes of  $k_x,k_y$ in the momentum integrals (see \ref{gap}).  Deforming the contour appropriately then the aforementioned-mentioned generalized Wick rotation  gives the following prescription for the momentum measure:
\bea
&&k_{x,y} \rightarrow i k_{2,3} \, ,
\nn \\
&&i \int d^3k \rightarrow  -i\int d^3k_{E_T} \, ,
\eea
 which, contrary to our standard  $3+1$ space-time, leaves the additional $i$ in front. 

In $E_T$ space  the singularity of the free propagator remains,   
for instance the free propagator of scalar particle is
\be
\frac{1}{p^{2}-m^2+\ep} \, ,
\ee 
with a positive square of the  three-momenta 
\be
p^2=p_1^2+p_2^2+p_3^2 \, ,
\ee
 thus formulation of the weak coupling (perturbation) theory, albeit possible, would not be more helpful then the standard approach (Wick rotation).

The advantage of the transformation to $E_T$ space becomes manifest, since the  fixed square Minkowski momentum $p^2=const$ hyperboloid with infinite surface is transformed into the finite  3dim-sphere in $E_T$ space. The Cartesian variables are related to the  spherical coordinates as usually:
\bea
&&k_3=k \cos\theta
\nn \\
&&k_1=k \sin\theta \cos \phi
\nn \\
&&k_2=k \sin\theta \sin \phi \, .
\eea

Making the aforementioned 2d Wick rotation, taking the Dirac trace on $\Sigma$ and integrating over the angles we get for the function $B$
\be \label{ET}
B(p)=m+i(2+\xi)\frac{e^2}{4\pi^2}\int_0^{\infty} dk
\frac{k}{p}\ln\left|\frac{k+p}{k-p}\right|S_s(k) \, ,
\ee
which is the same equation we derived directly in Minkowski space (\ref{ET1}).

Making the trace  $ Tr \not p\Sigma/(4p^2)$ and integrating over the angle we get for the renormalization wave function in $E_T$ space
\bea \label{ET2}
A(p)&=&1+i\frac{e^2}{4\pi^2}\int_0^{\infty} d k
\frac{k^2}{p^2} S_v(k^2) \left[-I+(1-\xi)I\right]
\nn \\
I&=&1+\frac{p^2+k^2}{2pl}\ln\left|\frac{k-p}{k+p}\right|
\eea
with the propagator function $S_v$ defined by (\ref{sv}).
The first term in the bracket $[]$ follows from the metric tensor while the second one proportional to
$(1-\xi)$  stems from the longitudinal part of gauge propagator.
We can see that like in the standard ES formulation  we get $A=1$ exactly  in quenched rainbow approximation in Landau gauge $\xi=0$.

\section{Numerical solutions for timelike $p^2$}

Before describing our result, at this place it is noteworthy to mention the study of complex singularities \cite{MARIS1995} which were found when  the momentum was continued to the complex plane. The author of \cite{MARIS1995} rotated integration contour and solve  QED2+1 SDEs system for complex variables selfconsistently (i.e. the same ray $p e^{i\phi}$ with  constant phase $\phi$ was considered for the external and the internal variables). The location of the singularity was obtained by extrapolation of the  complex solution, since at near vicinity of complex pole the numerical  procedure was unstable. Recall here, the phase $\pi/2$ would transform standard Euclidean SDE to the Temporal space used here, however this solution was never achieved because of singularity found. As  will be shown,  the solution become stable  again there,  since  we are enough far from the singularity. Actually, we have found the equation stable even when the pole is expected near real the timelike axis, since in our case the stability is guaranteed by a non-zero value of current fermion mass $m$ in this case.  Although we are not really interested in the location of the complex plane we can refer the result for the case $e^2>>m$ (only $m=0$ case were studied for all the all truncations of SDEs system in \cite{MARIS1995}). The observed location  of singularity is characterized by $\phi\simeq \pi/4$ and $|p|\simeq M(0)$ where $M(0)$ is the dynamical infrared mass. Apparently, the author did not continue the solution further to the timelike axis, because of crossing the singularity, however as we opposed above,  doing opposite could be the correct procedure. One should start from the Temporal Euclidean solution, which being identified with the physical timelike Minkowski solution, the continuation  to the spacelike regime could be performed afterwords (including the residua of the complex pole found).

In this paper we present the solution of the  mass function $M=B$ in the ladder approximation in Landau gauge in (\ref{ET}) space, we do not perform the  continuation to the spacelike axis, which remains to be done in the future. We will consider the nonzero Lagrangian mass  $m$ and interaction strength characterized by  charge $e$.
QED2+1 is  a superrenormalizable theory and  as we have neglected the photon polarization
it turns to be completely ultraviolet finite and  no renormalization is required at all.
 We assume that the imaginary part of the mass function is dynamically generated and to get the  numerical solution  we split the SDE (\ref{ET}) to the coupled equations for the real and imaginary part of $B$ and solve these  two coupled 
integral equations simultaneously  by the method of  iterations.

The complex mass generation observed herein is not only the simplest and most straightforward definition of quark confinement, but  it has an advantageous technical consequence. The associated absence of the real singularity at $p^2$ axis ensures that, for all $p^2$, the integrand is regular and hence the integral of SDE can be evaluated using straightforward Gaussian quadrature technique, i.e. there are no endpoint or pinch singularities on integration axis,  except  when the coupling constant is very small when compared to the bare mass $m$. In the later case the quantum corrections become negligible and the real branch point close to $m$ is restored and heavy electron (or weakly interacting) QED2+1 is an unconfined theory again.

To characterize complex mass dynamical generation it convenient to introduce dimensionless parameter
\be
\kappa=\frac{e^2}{2m}.
\ee
We set up the scale by taking $m=1$ at any units. Pure dynamical chiral symmetry breaking $(m=0)$ is naturally achieved  at large  $\kappa$ limit.

The phase $\phi_M$ of the complex mass function defined by $M=|M|e^{i\phi_M}$
is shown in Fig.1. for a various value of $\kappa$.  For very large $\kappa$ we get the dynamical chiral symmetry breaking in which case the obtained infrared phase  is $\phi_M(p^2=0,\kappa=\infty)=87.5^o$ while more interestingly it vanishes for very small $\kappa$. There is no confinement of fermion at all bellow some "critical" value of $\kappa$ , especially one can observe $\phi_M(p^2=0,\kappa<0.0191\pm 0.0001)=0$ for one flavor ladder QED2+1. The absolute value of $M$ is displayed in Fig.2. for the various value of the coupling $ \kappa$. As $\kappa$ decreases the expected complex singularities gradually moves from complex plane to the real axis and the function develops expected threshold enhancement, however here it is much smaller when compared to QED3+1.

\begin{figure}
\centerline{\epsfig{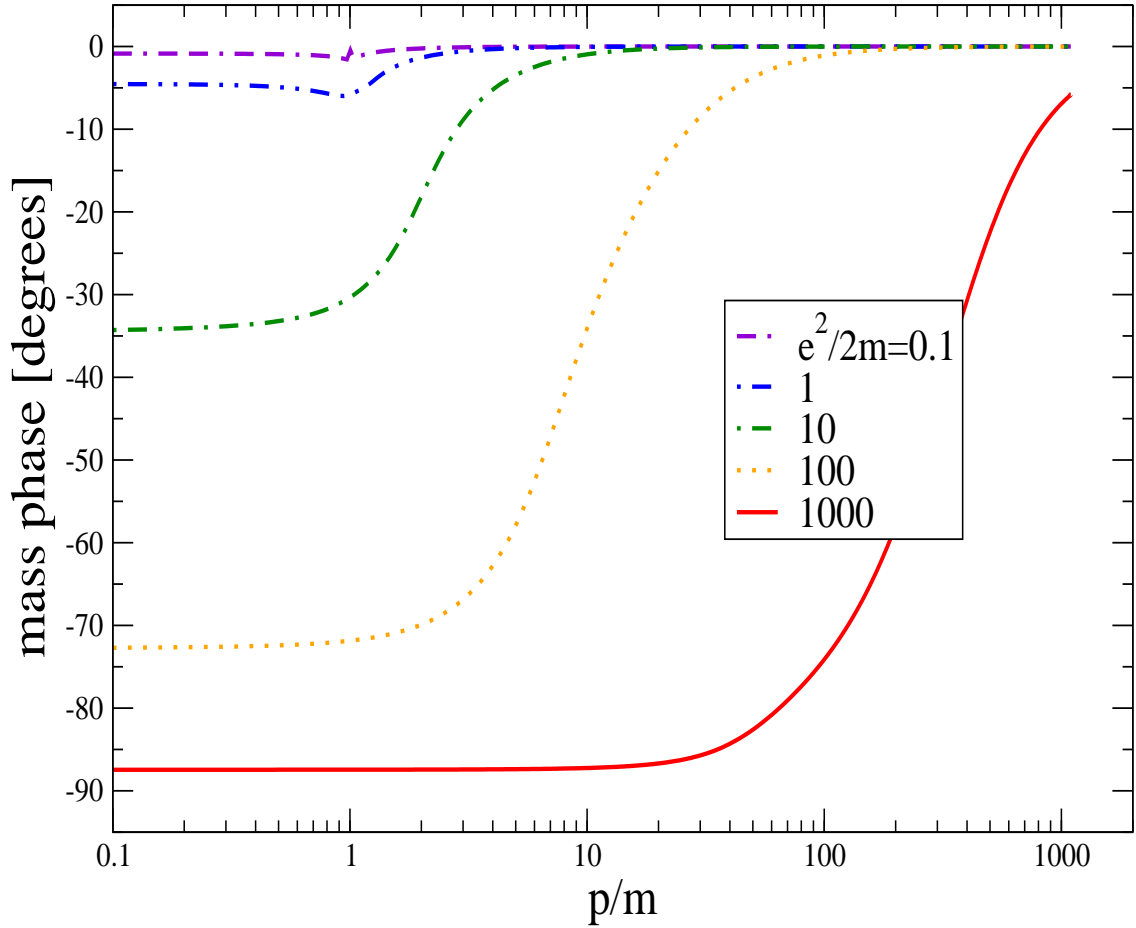}}
\caption[caption]{Phase $\phi $ of the dynamical mass function $M=|M|e^{i\phi}$ of electron living in 2+1 dimensions for different $\kappa$, scale is $m=1$.} \label{uhlyqed}
\end{figure}

\begin{figure}
\centerline{\epsfig{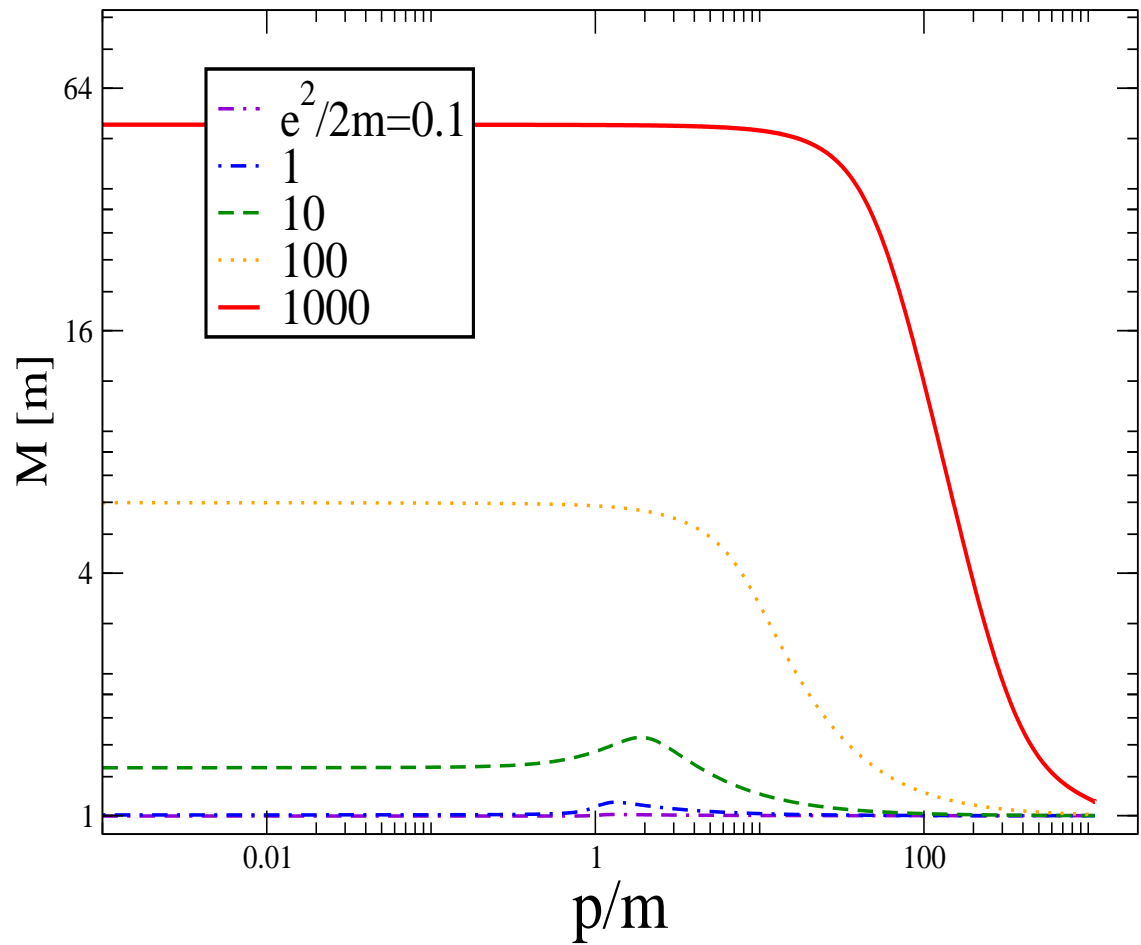}}
\caption[caption]{Magnitude $|M|$ of the running mass $M=|M|e^{i\phi}$ of electron living in 2+1 dimensions for different $\kappa$, scale is $m=1$.} \label{velqed}
\end{figure}

\section{Summary and conclusions}

We have presented the first analysis of the  electron gap equation  
in  Temporal Euclidean and Minkowski space. The main result, although based on the simple
ladder approximation in given gauge, is the proof of the exact equivalence between  the theories defined in Minkowski 2+1 and 3D Temporal Euclidean space. No similar is known about  the standard Euclidean formulation and its relation with spacelike  subspace of  Minkowski space.

The  dynamical generation of   imaginary part of the fermion mass can lead to the absence of real pole, providing simple scenario of confinement: there are no freely moving fermions if $\kappa=e^2/2m$ is not considerably  smaller then unity.
Minkowski QED2+1 has been shown to exhibit spontaneous chiral symmetry breaking -the mass function  has nontrivial solution for $m=0$ and it is confining if the electrons are enough light. We expect robust quantiative changes when the polarization effect is taken account.

The Temporal Euclidean space  introduced recently and used here opens up a variety of questions. If there is any, what is the relation between standard Euclidean (e.g. lattice) formulation with the original Minkowski space confining theory? We clearly argue and explain in details in the Section III that the Wick rotation -the well known calculational trick in quantum theory- is based on an unjustified assumption for QED2+1. 
More interestingly, what are the outcomes when the same ideology is applied to the strong coupling 
even dimensional theories, e.g. to the most successful theory of the strong interaction in the nature: QCD \cite{SABA2009}?

\begin{center}{\large \bf  Acknowledgments}\end{center}

I would like to V. Gusynin for useful comments. This work has been financially supported by the FCT Portugal and CFTP IST Lisbon.


\end{document}